\documentclass[runningheads]{llncs}

\usepackage{graphicx}
\usepackage{hyperref}

\usepackage{multirow}
\usepackage{longtable}
\usepackage{enumitem}
\usepackage{array}
\usepackage[dvipsnames]{xcolor}

\usepackage{tabularx}
    \newcolumntype{L}{>{\raggedright\arraybackslash}X}
    \newcolumntype{R}{>{\raggedleft\arraybackslash}X}
    \newcolumntype{b}{>{\hsize=.55\hsize\raggedright\arraybackslash}X}
    \newcolumntype{c}{>{\hsize=.55\hsize\centering\arraybackslash}X}
    \newcolumntype{s}{>{\hsize=.55\hsize\raggedleft\arraybackslash}X}

\usepackage{blindtext}
\usepackage[most]{tcolorbox} 
\definecolor{block-gray}{gray}{0.95}

\usepackage{caption} 
\newtcolorbox{zitat}[2][]{%
    colback=block-gray,
    grow to right by=-10mm,
    grow to left by=-10mm, 
    boxrule=0pt,
    boxsep=0pt,
    breakable,
    enhanced jigsaw,
    borderline west={4pt}{0pt}{gray},
    title={#2\par},
    colbacktitle={block-gray},
    coltitle={black},
    fonttitle={\large\bfseries},
    attach title to upper={},
    #1,
}

\usepackage{tikz}
\usepackage{amsmath}

\usetikzlibrary{positioning,fit}

\begin{document}
%
%\title{Uncovering the Hidden World of User Interaction Data Collection in Apps}
\title{Analytics for ``interaction with the service'': Surreptitious Collection of User Interaction Data}
\titlerunning{Surreptitious Collection of User Interaction Data}
\author{Feiyang Tang, Bjarte M. \O stvold}
\authorrunning{}
\institute{Norwegian Computing Center, Oslo, Norway \\
\email{\{feiyang, bjarte\}@nr.no}}
\maketitle              % typeset the header of the contribution
\begin{abstract}
% Problem
The rise of mobile apps has brought greater convenience and customization for users. However, many apps use analytics services to collect a wide range of user interaction data purportedly to improve their service, while presenting app users with vague or incomplete information about this collection in their privacy policies. Typically, such policies neglect to describe all types of user interaction data or how the data is collected.
% Why problem
User interaction data is not directly regulated by privacy legislation such as the GDPR. However, the extent and hidden nature of its collection means both that apps are walking a legal tightrope and that users' trust is at risk.

% Startling sentence
To facilitate transparency and comparison, and based on common phrases used in published privacy policies and Android documentation, we make a standardized collection-claim template.
%For the top 10 apps, we compare in detail their collection claims, derived from their privacy policies, to their actual collection, and find that all their claims are incomplete.
Based on static analysis of actual data collection implementations, we compare the privacy policy claims of the top 10 apps to fact-checked collection claims. Our findings reveal that all the claims made by these apps are incomplete.
% Implication
By providing a standardized way of describing user interaction data collection in mobile apps and comparing actual collection practices to privacy policies, this work aims to increase transparency and establish trust between app developers and users.

\keywords{Mobile Apps \and User Interaction Data \and User Trust \and Static Analysis\and Privacy Policy}
\end{abstract}

%-------------------------------------------------------------------------------
\section{Introduction}\label{Sec:intro}
%-------------------------------------------------------------------------------
Mobile apps have enabled us to interact with technology in new ways, allowing app developers to collect large amounts of user interaction data with the help of analytics services such as AppsFlyer, Flurry, and Firebase Analytics. 
%User interaction data encompasses the actions taken by users, such as tapping buttons, scrolling through pages, and watching videos. The nature and extent of this data collection will be a surprise to many users since apps typically do this without informing them. 
%The lack of transparency in data collection is an issue, particularly when it comes to the vagueness and incompleteness of privacy policies, leaving users unsure of what data is being collected and how it is being used. 
%This can lead to a sense of mistrust among users, potentially resulting in a loss of confidence in mobile apps~\cite{Cysneiros2009AnIA}. 
User interaction data includes actions taken by users, such as tapping buttons, scrolling through pages, and watching videos. Unfortunately, in their privacy policies apps often use vague phrases like ``user's interaction with the service'' to describe data collection without providing any further details. 
This lack of granularity exacerbates the problem of transparency, leaving users unsure of the scope and nature of the data being collected and the ways in which it is being used.
As a result, users may feel a sense of mistrust towards apps, leading to a loss of confidence in their use.

The relationship between transparency of data collection and user trust is crucial~\cite{Cysneiros2009AnIA}. 
Without transparency, users cannot make informed decisions about what data they share and how it is used~\cite{morey2015customer}, which can result in a sense of invasion of privacy. 
Additionally, data collected without proper transparency and control can be used for various purposes, including targeted advertising and user profiling, further eroding user trust.

An example is the Yr app, which is the most popular weather app in Norway, developed by the Norwegian Broadcasting Corporation (NRK). The app collects user interaction data to understand which features are frequently used, such as when users set their location to get localized forecasts.
By analyzing this data, the app could potentially infer that a user is a frequent traveler, has family living in different locations, or is interested in travel-related products or services. 
Such profiling and inference could be used to target the user with location-based advertising, without the user's knowledge.
However, the app's privacy policy\footnote{\url{https://hjelp.yr.no/hc/en-us/articles/360003337614-Privacy-policy}} is vague regarding the collection of user interaction data, as it lacks specific details on the types of data collected, as displayed below in the gray box.

\begin{center}
\resizebox{\textwidth}{!}{% <------ Don't forget this %
\begin{zitat}{Privacy Policy of Yr}
$\cdots$

We use different tools to track the use on our app and website. This information gives us valuable information such as most popular pages and on what times Yr is being used the most. No information that can identify persons are available for Yr.

$\cdots$

\end{zitat}% <------ Don't forget this %
}
\end{center}

Our reading of NRK's privacy policy\footnote{\url{https://info.nrk.no/personvernerklaering/}} did not yield any specific information on Yr's data collection practices, as the policy is primarily focused on NRK's news services and their ``interaction with the services'' collection practices. 
This lack of transparency regarding the data collection practices of the Yr app is concerning, as it can erode user trust in the app and NRK as a whole.

Recent studies have shown that even seemingly innocuous user interaction data can reveal sensitive information about individual users. For instance, emoji use and pages visited can be used to infer a user's preferred pool and political orientation, respectively \cite{gadotti2022pool}. Additionally, mobile biometric data associated with keystrokes and touchscreen gestures can be used to estimate soft attributes like age, gender, and operating hand \cite{7736910,jain2019gender}. 
These findings highlight the potential risks associated with collecting user interaction data, which can be used to infer sensitive information about individual users and subsequently be used for user profiling.
Moreover, interaction data is not directly linked to identifying an individual and therefore may not be covered by major data protection regulations such as the GDPR.
%These findings emphasize the need for more research on how to apply privacy protections for multiple collections.

Most current research studying the privacy implications of analytic services has focused on determining whether personally identifiable information (PII) is being collected and transmitted to external analytics services~\cite{9284006,8660581}. Studies have also examined log data to understand user behavior~\cite{dumais2014understanding}, and high-level analyses of user behavior data collection in mobile apps have been conducted~\cite{verkasalo2010analysis}.

%This motivated us to combine static analysis with the study of user interaction data collection claims in privacy policies. By doing so, we aimed to increase transparency and build user trust by providing a clearer understanding of how user data is being collected and used in mobile apps.
%Very little research has been done on how user interaction data itself is collected within mobile apps. 

\subsection{Research Questions}
%Building upon the introduction, the aim of this paper is to provide a more standardized and structured way of describing user interaction data collection in mobile apps, as well as ensuring that the claims made by mobile apps are accurate and transparent.
Building upon the introduction, the aim of this paper is to provide a more standardized and structured way of describing user interaction data and how is collected and used in mobile apps and furthermore to increase transparency by comparing the claims made by mobile apps in their privacy policies to their actual implementation of collection. This approach is motivated by the need to build user trust. % by providing a clearer understanding of how user data is being collected and used in mobile apps.
In order to achieve this aim, we have devised a set of research questions:

\begin{enumerate}[font={\bfseries},label=RQ\arabic*, ref=\arabic*]
   \item \label{rq:rq1} What do apps say about their collection of user interaction data in their privacy policies? We refer to this as their \emph{privacy policy collection claims}.
   \item \label{rq:rq2} What types of user interaction data do apps actually collect in their implementations, and what are the means used for this collection?
   \item \label{rq:rq3} How do the collection claims from the policies match up with the realities of the implementations? We shall refer to this as \emph{checking the claims}.
\end{enumerate}

\subsection{Contributions}
In this paper, we made several contributions to the understanding of user interaction data collection practices in mobile apps:
\begin{itemize}
    \item We analyze a corpus of privacy policies in mobile apps to identify the common phrasing used to describe the collection of user interaction data. 
    %Our analysis revealed that the majority of these policies lacked completeness in describing such collection practices. 
% (Section~\ref{sec:claim})
%    \item 
    We then create a \emph{standardized collection claim template} for user interaction data based on the common phrasing and a vocabulary derived from Android documentation (Section~\ref{sec:claim}).
    \item We use static analysis to extract collection evidence from Android apps, identifying data types, relevant code and means of collection in layout files and bytecode (Section~\ref{Sec:analysis}).
    \item %To evaluate the completeness and consistency of privacy policies, we fact-checked 10 top downloaded Android apps from top 10 popular cateories and provided an overview of user interaction data collection practices in 100 popular apps on Google Play 
    We provide an overview of user interaction data collection practices in 100 popular apps on Google Play, representing the top 10 popular categories. From this sample, we selected the most popular app from each category, 10 apps in total, and conducted a fact-check of their privacy policy collection claims using collection evidence (Section~\ref{sec:finding}).
\end{itemize}

%-------------------------------------------------------------------------------
\section{Background}
%-------------------------------------------------------------------------------
Before delving into the specifics of our investigation, it is important to establish a foundation of knowledge on the topic of user interaction data collection in Android apps. 
This involves examining the human aspects of this issue, understanding the basic components of an Android app, and exploring common reverse engineering techniques used to analyze app behavior. 
Additionally, we will discuss what user interaction data entails, and how apps collect this data through various means such as logging interactions, timing interactions, and recording series of interactions. 

\subsection{User Trust and Transparency}

Ensuring user trust and transparency in mobile app data collection is crucial. Clear, transparent, and reversible procedures can help establish user confidence and control over their data, leading to better user experiences and app adoption. Conversely, a lack of transparency can erode trust, create privacy concerns, and lead to user dissatisfaction or abandonment of the app. 
%Studies have shown that higher transparency is linked to stronger participant trust~\cite{fischer2016transparency}, while low transparency can hinder adoption~\cite{vorm2022integrating}. Prioritizing transparency and user control can help establish trust and confidence, resulting in a better user experience and increased adoption.
Research has shown that increased transparency is positively correlated with higher levels of trust among participants~\cite{fischer2016transparency}, whereas low transparency can impede the adoption of mobile apps~\cite{vorm2022integrating}. Given the importance of building trust and confidence among users, it is in the interest of app makers to prioritize transparency and user control. Such efforts can ultimately result in a better user experience and increased uptake of their apps.

\subsection{Android Reverse Engineering and Analytics Services}

APK files contain all the resources and bytecode necessary for Android apps to function, with several activities providing the user interface. Tools like apktool\footnote{\url{https://ibotpeaches.github.io/Apktool}} and Dex2Jar\footnote{\url{https://github.com/pxb1988/dex2jar}} can be used to decompile APKs and analyze analytics libraries.

Analytics services like Firebase Analytics and Flurry Analytics enable developers to gather data on user behavior, engagement, and preferences. 
However, these services have raised concerns about data protection as they often automatically collect user data, thereby raising privacy concerns. Thus far, countries such as France, Italy, Austria, Denmark, and Norway have clearly stated that the use of Google Analytics violates GDPR~\cite{SeveralE79:online}.
Android apps can use these services by directly invoking third-party APIs or customizing their own analytics service by extending these APIs. The first approach involves calling third-party API methods directly in activities to log user engagement events, while the second approach enables developers to tailor data collection to their specific needs.

%-------------------------------------------------------------------------------
\section{Making sense of App Policies about Data Collection}\label{sec:claim}
%-------------------------------------------------------------------------------
To address \textbf{RQ\ref{rq:rq1}}, we investigate how mobile apps disclose their collection practices for user interaction data in their privacy policies. 
To make sense of app policies we propose a standardized template of ``data collection \emph{claims}'' as a single sentence in a restricted vocabulary. This sentence gives the essence of interaction data collection: the specific data types collected and the means of collection.

\subsection{Collection Vocabulary}\label{Sec:techvoc}
Our restricted collection vocabulary was developed by analyzing the Android system implementation documentation, as well as the APIs of the top 20 analytic services for Android apps listed on AppBrain~\cite{Androida43:online}.%, which provides the source of information about Android apps.

\subsubsection{Terms for Types of User Interaction Data}
The user interface of an Android app collects a variety of data types, such as touch events, sensor data, and text input.  Based on a manual inspection of every single type of Android UI widget, we identified the following six types of interaction data and named them:
%These data types result from users' interaction with the app.% and can be used to improve the user experience and inform app development.
%We use the following terms to describe the identified types of interaction data that are collected through the app user interface.
\begin{itemize}
    \item \emph{App presentation interaction data}: This data arise from the consumption of content provided by the app. For example, the user plays a certain video for a period of time, spends minutes reading one specific page of the news. These interactions are often recorded by a logging system to keep track of the user's consumption habits. %For example, the app might keep a record of the specific news pages a user has opened and how long they spent on that page.
    \item \emph{Binary interaction data}: This data arise from discrete user actions, such as tapping on a button or icon, or selecting a checkbox. %Apps can log these interactions to track when and how often the user performs these actions.
    \item \emph{Categorical interaction data}: This data arise from a selection from a set of predefined options or categories, such as choosing a value from a dropdown menu, selecting a radio button, or rating a product. %These interactions are often logged by the app to track the user's preferences and choices.
    \item \emph{User input interactions data}: This data arise from user input through an on-screen keyboard or another input method, such as entering text or numbers into a form field, or using voice input to perform a search or command. %These interactions can provide valuable insights into the user's interests and needs.
    \item \emph{Gesture interactions data}: This data arise from gesture inputs and smooth and continuous movements of the user's finger on the screen, such as scrolling through a list, swiping left or right, pinching or zooming, or shaking the device. %These interactions can be used to provide additional functionality or to provide a more intuitive user experience.
    \item \emph{Composite gestures data}: This data arise from a combination of multiple gestures, such as tapping and holding, double tag, or drag and drop. %These interactions can provide even more advanced functionality or a more complex user experience.
\end{itemize}

\subsubsection{Terms for Means of Collection}
%We developed a set of standardized vocabularies to describe the means of user interaction data collection in Android apps. 
%These vocabularies provide clear and consistent terminology for describing how apps collect user interaction data and through what means.
%These types of interactions can be used in various ways by apps to collect data about the user's behavior and preferences.
We use the following terms to describe the means of user interaction data collection.
\begin{itemize}
    \item \emph{Frequency}: This type of collection involves logging the frequency of the occurrence of a particular interaction. For example, an app might log the number of times a user taps on a specific button or selects a certain option from a drop-down menu. %This data can provide detailed insights into the user's preferences and habits.
    \item \emph{Duration}: This type of collection involves tracking the time a user spends engaging in a particular interaction. For example, an app might log the amount of time a user spends watching a particular video or reading a specific article. %This data can provide insights into the user's level of interest in a particular topic and can be used for market research purposes.
    \item \emph{Motion details}: This type of collection involves monitoring the specific details of a user's interaction, such as the speed, direction, or angle of their finger movements on the screen. This type of data can be collected for interactions such as scrolling, swiping, or dragging. %Motion details can be used to infer the user's preferences and interests, such as their preferred method of navigation within an app or their level of engagement with particular elements on the screen.
\end{itemize}

\subsection{From Policies to Standardized Collection Claims}\label{Sec:pp}
%The privacy policy claims involve providing descriptive explanations for each data collection. 
%These descriptions are obtained from the 
We start by examining the privacy policies of publicly available mobile apps. 
%To obtain the relevant privacy policies,
We used the APP-350 Corpus~\cite{story2019natural}, which consists of 350 mobile app privacy policies that are annotated with privacy practices. 
Since APP-350 focuses on identifying personally identifiable information (PII) related sentences from the privacy policies, we only used their raw privacy policy HTML files for our analysis.

\subsubsection{Identifying Relevant Policy Parts}

To identify sentences related to user interaction data collection in privacy policies, a simple pre-trained language model was adopted. The model processes HTML files of privacy policies and extracts sentences that contain specific keywords and their synonyms. Natural language processing is performed using the spaCy~\cite{spacy2} library with the \verb|en_core_web_sm| model. This model is pre-trained on web text, which includes web forums, web pages, and Wikipedia. It can identify named entities, parts of speech, dependency parsing, and more. The WordNet module from the Natural Language Toolkit (NLTK~\cite{bird2009natural}) is also used to find synonyms for the extracted keywords.

To validate the effectiveness of the model, we randomly selected 50 privacy policies from the APP-350 dataset and manually annotated them to identify sentences containing relevant information, the verbs used to describe data collection (e.g., ``collect'', ``track''), and the terms used to describe user interaction data (e.g., ``usage of the app'', ``interaction with the service'').
By identifying the verb and the terms most commonly used to describe user interaction data
%in privacy policies to describe the collection of user interaction data, 
we ensure that our privacy policy claim vocabulary is consistent with prevailing industry practices.% and accurately reflects how user interaction data is collected.

The model successfully identified sentences related to user interaction data collection in 37 of the 38 files that contained such sentences, using keywords such as interaction, usage, statistics, experience, and analytics. We confirmed that the 37 identified files contained relevant sentences. Identifying the verbs used to describe data collection was more challenging, with 92\% recall but only 84\% precision
%\footnote{Recall is calculated as TP / (TP + FN), while precision is calculated as TP / (TP + FP), where TP is true positives, FN is false negatives, FP is false positives, and TN is true negatives.} 
due to similar verbs appearing in sentences that were not related to the context.

%However, the small number of false positives did not affect the analysis since the top frequently appearing terms were manually checked in the end. 
%After running the model on the 350 privacy policies and performing manual checks, we identified common phrases used in these policies to describe the collection of user interaction data from the 1411 identified sentences, as shown below in Table~\ref{Tab:keyword1} and Table~\ref{Tab:keyword2}.
After running the model on the 350 privacy policies and performing manual checks to eliminate any false positives, we identified common phrases used in these policies to describe the collection of user interaction data. From the 1411 identified sentences, we selected only the relevant verbs and nouns to compile the list of keywords shown below in Table~\ref{Tab:keyword1} and Table~\ref{Tab:keyword2}.

\begin{table}[htbp]
\centering
\resizebox{.95\textwidth}{!}{%
    \begin{minipage}{.5\textwidth}
      \caption{The top five most frequent terms used to describe user interaction data in the APP-350 corpus}
        \label{Tab:keyword1}
      \centering
        \begin{tabular}{lr}
            %\hline
            \textbf{Term} & \textbf{Count}\\\hline\hline
            interact($\sim$ion,$\sim$ing) with service/app & 1049 \\
            analytic(s) & 886 \\
            us($\sim$age, $\sim$ing) of service/app & 397 \\
            statistic(s) & 315 \\
            input(s) of user & 173 \\
            \hline
        \end{tabular}
    \end{minipage}%
    \hspace{1cm}
    \begin{minipage}{.5\textwidth}
      \centering
        \caption{The top five most frequent verbs used to describe such collection in the APP-350 corpus}
        \label{Tab:keyword2}
        \begin{tabular}{lr}
            %\hline
            \textbf{Verb} & \textbf{Count}\\\hline\hline
            collect & 1386 \\
            track & 548 \\
            use & 202 \\
            log & 86 \\
            gather & 46 \\
            \hline
        \end{tabular}
    \end{minipage} %
    }
\end{table}

%In privacy policies, it is common to use convoluted language to describe how user data is collected. To standardize this language, we utilized the most frequently used verb, ``collect'' and the most frequently used noun phrase ``interaction data'' and created a straightforward structure to describe the policy's claim for the collection of user interaction data.
\subsubsection{A Template for Standardized Collection Claims}
In privacy policies, it is common for apps to use convoluted language to describe how user data is collected. 
To make these collection claims in privacy policy easier to read and compare across different apps, we created a standardized template that utilized the most frequently used verb, ``collect'', and the most frequently used noun phrase, ``user interaction data''.
The resulting structure is as follows:

\begin{center}
\resizebox{\textwidth}{!}{% <------ Don't forget this %
\begin{zitat}{Template for Standardized Collection Claims}
We collect the following types of user interaction data: $\langle$\textit{types of data collected}$\rangle$, along with their $\langle$\textit{means of collection}$\rangle$.\footnote{Refer to the claim vocabularies in Section~\ref{Sec:techvoc}}
\end{zitat}% <------ Don't forget this %
}
\end{center}

%By presenting privacy policies using this template, we can make it easier to compare the collection claims made by different apps. 
This standardized collection claim template can be combined with the collection evidence gathered through static analysis to check and the accuracy of privacy policy collection claims made by various apps. Also, the standardized language facilitates transparency and comparison between policies. 
We return to the subject of checking policy claims in Section~\ref{sec:finding}.

\section{Data Collection Evidence}\label{Sec:analysis}
To understand what user interaction data apps actually collect (\textbf{RQ\ref{rq:rq2}}) and to check the apps' privacy policy collection claims (\textbf{RQ\ref{rq:rq3}}) we analyze the Android apps' implementations:
Firstly, we identified the data collection methods used in the app package (APK). 
Next, we extracted the relevant evidence from the app and mapped it to our collection vocabulary in Section~\ref{Sec:techvoc}. 
Lastly, we compared the evidence extracted from the app with the claims made in the privacy policy to evaluate the level of agreement between the two, which we refer to as claim checking. 

\subsection{Identifying Data Collection Methods}
%The first step was to identify the data collection methods used by the analytics services in the APK. 
Data collection methods (DCMs) are methods defined by analytics services, such as Firebase Analytics, that allow app developers to log user interaction data. 
DCMs provide a standardized way for app developers to collect user interaction data and track app usage in order to analyze and understand user behavior.

For example, the Firebase Analytics API provides the \texttt{logEvent()} method to log user events, such as button clicks or screen views.
Suppose we have a button \texttt{myButton} in the app's UI, and we want to track when the user clicks on it. 
We can do this using Firebase Analytics by adding the following code to the button's \texttt{OnClickListener}:
\begin{center}
\begin{small}
    \begin{verbatim}
        myButton.setOnClickListener(new View.OnClickListener() {
            public void onClick(View v) {
                FirebaseAnalytics.getInstance(this).
                    logEvent("button_click", null); }
        });
    \end{verbatim}
\end{small}
\end{center}
Here \texttt{FirebaseAnalytics.getInstance(this)} returns an instance of the Firebase Analytics object, and \texttt{logEvent("button\_click", null)} collects the button click interaction data with the string \texttt{"button\_click"} to Firebase Analytics. 
%The null parameter indicates that we are not passing any additional data with the event.
%The \texttt{logEvent()} method is a DCM defined by Firebase Analytics to track user interactions in Android apps. 
%Other analytics services have similar methods for logging events, such as Google Analytics' \texttt{send()} method.

To determine how Android apps use analytics services, we identified DCMs from the top 20 Android analytic services, cf.\ Section~\ref{Sec:techvoc}.
Matching the full signature of these methods in bytecode allows us to find direct invocations to analytics services. %, which is the most common and simple approach. 
However, some apps use customized analytics services to do a more fine-grained collection, such as collecting motion details and duration. To do this, the apps implement their own analytics classes by extending the analytic services.
To identify customized analytics, we use static analysis to identify the classes that invoke external DCMs. 
We then check whether these classes are invoked in any of the app's declared activities. 
If they are, we mark these classes as customized analytics services classes. 

%his method allows us to provide insights into how user interaction data is being collected by different Android apps, including the use of customized analytics services. It also allows us to better understand how user interaction data is being processed and analyzed by these analytics services, which can have significant implications for user privacy and trust.

\subsection{Extracting Collection Evidence}
Next, we extracted evidence of actual data collection from the APK. 
Specifically, we analyze three types of information: (1) invocations to analytics services that logged user interaction data collection, (2) associated UI widgets, and (3) the callbacks triggered by registered listeners on these UI widgets.

We utilized static analysis with FlowDroid~\cite{arzt2014flowdroid} to associate DCM invocations with callbacks, listeners, and activities in the bytecode. 
We then compared the layout IDs of the associated UI widgets defined in the layout XML files to identify the relevant collection data types and means. 
%By doing so, we were able to fit the extracted evidence into our claim vocabularies and generate a standardized privacy policy claim.
%By doing so, we were able to fit the extracted evidence into our collection vocabulary and compare it against our standardized privacy policy claim, enabling us to identify inconsistencies and incomplete claims related to user interaction data collection.

\begin{figure}[htbp]
\centering
    \begin{tikzpicture}
    \begin{scope}
    [every node/.style = {shape=rectangle, rounded corners, draw, align=center, ->}]
    \node(A)                    {App};
    \node(B)       [below=of A] {Layout files};
    \node(C)       [left=of B]  {Activity};
    \node(D)       [below=of B] {UI widget};
    \node(E)       [below=of C,left=of D] {Listener};
    \node(F)       [left=of E]  {Callback};
    \node(G)       [above=of F,left=of C] {DCM \\ Invocation};
    \end{scope}
    
    \draw  (A) to node [right, near end] {\tiny M} node [right, near start] {\tiny 1} (B);
    \draw  (A) to node [left, near end] {\tiny M} node [left, near start] {\tiny 1} (C);
    \draw  (B) to node [right, near end] {\tiny M} node [right, near start] {\tiny 1} (D);
    \draw  (C) to node [right, near end] {\tiny M} node [right, near start] {\tiny 1} (D);
    \draw  (C) to node [left, near end] {\tiny M} node [left, near start] {\tiny 1} (E);
    \draw  (C) to node [left, near end] {\tiny M} node [left, near start] {\tiny 1} (F);
    \draw  (C) to node [above, near end] {\tiny *} node [above, near start] {\tiny 1} (G);
    \draw  (F) to node [left, near end] {\tiny 1} node [left, near start] {\tiny *} (G);
    \draw  (E) to node [above, near end] {\tiny M} node [above, near start] {\tiny 1} (F);
    \draw  (E) to node [above, near end] {\tiny 1} node [above, near start] {\tiny 1} (D);

    \node[draw=none, rounded corners=0.3cm, fill opacity=0.1, fill=blue, inner sep=10pt,label={[blue!80]above:Means of collection}, fit=(C)(E)(F)(G)] (M) {};
    \node[draw=none, rounded corners=0.3cm, fill opacity=0.1, fill=red, inner sep=10pt,label={[red!80]below:User interaction data types}, fit=(B)(D)] (U) {};
    \end{tikzpicture}
\caption{Relationships between different parts of the extracted collection evidence in an Android app} 
\label{fig:relationship}
\end{figure}
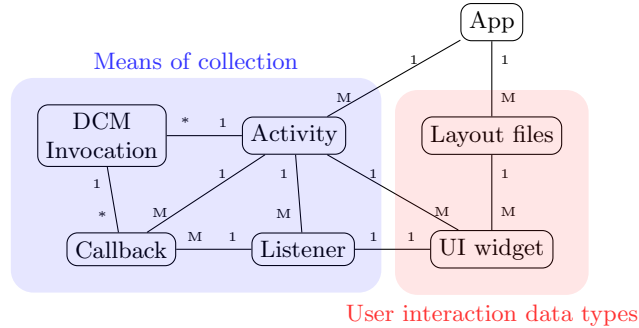

The relationships between different parts of the extracted collection evidence in an Android app are shown in Fig.~\ref{fig:relationship}.
The UI-related parts, such as layout files and defined UI widgets, provide information on the types of user interaction data (red section), while the bytecode provides details on the means of collection (blue section)\footnote{Note: The figure notation is as follows: 1-M means one-to-many, 1- means one-to-any (zero or more), and 1-1 means one-to-one.}.

We return to the Yr weather app, the example app from Section~\ref{Sec:intro}. Based on the collection evince extracted from Yr's bytecode and layout files, we discovered that it collects detailed user interaction data using various types of UI widgets such as SearchView and Textfield. This data collection is linked to features such as changes in location, enabling forecast summary notifications, and opening the forecast graph. Building on this finding from static analysis, we propose the following more specific checked standardized collection claim:

\begin{center}
\resizebox{\textwidth}{!}{% <------ Don't forget this %
\begin{zitat}{Checked Standardized Collection Claim for Yr}
$\cdots$

We collect the following types of user interaction data: \textit{app presentation, binary and categorical interactions, and user input interactions}, along with their \textit{frequency}.

$\cdots$
\end{zitat}% <------ Don't forget this %
}
\end{center}

\section{Findings}\label{sec:finding}

To address \textbf{RQ\ref{rq:rq1}}, we conducted a manual inspection of 1411 sentences that described user interaction data collection in all 350 privacy policies within the APP-350 corpus, as outlined in Section~\ref{Sec:pp}.

We examine whether the sentences in one privacy policy provide clear descriptions of the types of user interaction data collected and the means of collection.

\begin{tcolorbox}
Our findings revealed that only 37\% of the identified sentences contained clear statements on both the data types and means of collection, while 41\% only discussed the means of collection and 22\% mentioned only the data types.
\end{tcolorbox}

Here are the relevant sentences from two policies in the corpus. DAMI\footnote{\url{https://play.google.com/store/apps/details?id=com.blappsta.damisch}} states: ``\textit{We may work with analytics companies to help us understand how the Applications are being used, such as the frequency and duration of usage.}'' Wish\footnote{\url{https://play.google.com/store/apps/details?id=com.contextlogic.wish}} states: ``\textit{We may collect different types of personal and other information based on how you interact with our products and services. Some examples include: Equipment, Performance, Websites Usage, Viewing and other Technical Information about your use of our network, services, products or websites}.''

DAMI's privacy policy only discloses the means of collection, such as the frequency and duration of usage, without clearly explaining which type of user interaction data is collected. In contrast, Wish's privacy policy does mention some specific types of data collected, such as equipment and performance data, but it is unclear about which means of collection are used. 
%These findings highlight the need for standardized and comprehensive collection vocabulary to improve privacy policy transparency and user understanding.

The majority of identified sentences discuss the means of collection rather than specific data types, suggesting that organizations use the tactic of avoiding or minimizing disclosures about the types of user interaction data they collect in order to collect more data than users are aware of or comfortable with. 
%It highlights the need for more transparent disclosures in privacy policies and the development of standardized collection vocabulary to improve transparency and increase user trust.

To investigate \textbf{RQ\ref{rq:rq2}}, we performed a static analysis on a sample of 100 free Android apps downloaded from the top 10 most popular categories on the German Google Play store\footnote{\url{https://play.google.com/store/apps?gl=DE}}, as identified by AppBrain.
In cases where the same app appeared in several categories, we moved to the next popular app in the second category to get a total of 100 distinct apps. 

\begin{table}[htbp]
\caption{Detailed analysis of the user interaction data collection for the top 100 Android apps.}
\label{Tab:appresults}
%\scriptsize
\centering
\resizebox{\textwidth}{!}{%
\renewcommand{\arraystretch}{1.5}
    \begin{tabularx}{\textwidth}{LLLss} 
    \textbf{UI type (types of interaction data)} &
      \textbf{Top 2 means of collection} &
      \textbf{Top 3 app categories} &
      \textbf{Percent collected} &
      \textbf{Average \# collected} \\\hline\hline
    View (Presentation)               & Frequency (100\%), Duration (52\%) & Entertainment, Shopping, Travel & 89\% & 12 \\
    Button (Binary)                   & Frequency (94\%), Motion (8\%)     & Social, Utility, Gaming         & 76\% & 26 \\
    Textfield (Input)                 & Frequency (100\%), Duration (4\%)  & Social, Shopping, Utility       & 63\% & 5  \\
    Checkbox \& Spinner (Categorical) & Frequency (97\%), Motion (16\%)    & Shopping, Travel, Utility       & 32\% & 7  \\
    GestureDetector (Gesture)         & Motion (94\%), Duration (40\%)     & Gaming, Entertainment, Social   & 16\% & 38 \\ \hline
    \end{tabularx}%
}
\end{table}

\begin{small}
\begin{tcolorbox}
%Our analysis revealed that the most commonly collected types of UI widgets in the top 100 Android apps were \textbf{Views, Buttons, TextFields, Spinners \& Checkboxes, and GestureDetector}. The most common means of collection was \textbf{frequency}. %Our findings also suggest a diverse range of data collection practices, with different types of user interaction data collected using various means of collection and across different types of applications.
Our analysis of the top 100 Android apps revealed that app developers placed a great deal of emphasis on understanding how frequently users interacted with different UI elements (which may correspond to different features or functionalities in the app), as frequency was the top means of collecting user interaction data across all UI types. We also found that the average number of interaction data collected varied significantly across different types of UI. It was also interesting to see that the high number of interaction data collected for the button UI type (also found in 76\% of the apps), indicated that understanding button usage was a particularly important metric for app developers.

\end{tcolorbox}
\end{small}

Table~\ref{Tab:appresults} compares collection practices across various app categories. 
Gaming apps collect a high percentage of Gesture data and send it to the analytics services later, likely due to the importance of tracking user finger motions for an optimal gaming experience. 
Entertainment, Shopping, and Travel apps tend to collect a high percentage of View data, indicating the importance of visual design customization and its consequences for the revenue stream of the app's service.
%and potentially allowing for the collection of user preferences on items. This could be attributed to the revenue benefits of customization in these types of apps.
Social and Utility apps concentrate on Button and Textfield data, reflecting their dependence on user input to perform actions and provide information. 
In summary, the result reveals differences in the app categories that prioritize understanding user interaction data depending on the UI type. 
This suggests that app developers may be focusing their efforts on understanding certain aspects of user behavior, such as how users view content, interact with buttons, or use gestures, depending on the specific goals and requirements of their app. 
Such insight could help app developers better tailor their data collection and analysis strategies to achieve their specific objectives.

To address \textbf{RQ\ref{rq:rq3}}, we manually inspected the privacy policy claims of the most popular app in each of the 10 categories on Google Play. We generated our checked collection claims by analyzing the actual data collection practices of each app and comparing them to the privacy policy claims published by the app.
Our checked collection claims are made by combining the evidence gathered through static analysis and the proposed standardized claim template.
The results are fact-check collection claims presented in Table~\ref{Tab:factcheck}.

\begin{tcolorbox}
Our study uncovered inconsistencies between the claims made in privacy policies and the actual data types and means of collection used by popular apps on Google Play. Many apps do not fully disclose the types of data collected or the means of collection, often using vague language such as ``collecting user interactions to improve the service''. 
\end{tcolorbox}

Notably, some apps that may be perceived as having questionable data collection practices, such as TikTok and Amazon Prime Video, actually provided more detailed information on the types of data collected and the means of collection used. TikTok and Duolingo even provided specific examples of their data collection practices.

However, we found that some apps from less controversial categories, such as the photography editing app Picsart and the payment platform PayPal, used opaque language in their privacy policies, leaving a large gap between their claims and our findings. The most extreme example was Booking.com, which extensively collects user interactions within the app, yet discloses almost no information in its privacy policy. These findings highlight the need for clearer and more comprehensive disclosures in privacy policies, particularly for apps that collect sensitive user data.

\begin{table*}[htbp]
\caption{Fact-checked data collection claims of the most popular app from each of the top 10 categories of Google Play. The red text indicates types of user interaction data missing from the existing privacy policy claims based on our checked standardized collection claims built on static analysis results, while the blue text indicates undisclosed means of collection.}
\label{Tab:factcheck}
\scriptsize
\centering
\resizebox{\textwidth}{!}{%
\renewcommand{\arraystretch}{1.5}
\begin{tabular}{p{0.34\textwidth}@{\hskip 0.2in}p{0.65\textwidth}}
  \textbf{Checked Collection Claim} &
  \textbf{Related Text in the Published Privacy Policy} \\\hline\hline
%TikTok &
  \textbf{[TikTok]} We collect the following types of user interaction data: app presentation, binary, {\color{red!60} categorical}, user input, gesture and composite gesture interactions, along with their frequency, duration and {\color{blue!60}motion details}. &
  \textbf{[TikTok]} We collect information about how you engage with the Platform, including information about the content you view, the duration and frequency of your use, your engagement with other users, your search history and your settings.\\
%SHEIN &
  \textbf{[SHEIN]} We collect the following types of user interaction data: app presentation, binary, categorical, {\color{red!60} user input interactions}, along with their {\color{blue!60} frequency and duration}. &
  \textbf{[SHEIN]} Data about how you engage with our Services, such as browsing, adding to your shopping cart, saving items, placing an order, and returns for market research, statistical analysis, and the display of personalized advertising based on your activity on our site and inferred interests; Collect your device information, and usage data on our website or app for fault analysis, troubleshooting, and system maintenance, as well as setting default options for you, such as language and currency. The display of information you choose to post on public areas of the Services, for example, a customer review. \\
%Booking.com &
  \textbf{[Booking.com]} We collect the following types of user interaction data: {\color{red!60}app presentation, binary, categorical, user input interactions}, along with their {\color{blue!60}frequency and duration}. &
  \textbf{[Booking.com]} We collect data that identifies the device, as well as data about your device-specific settings and characteristics, app crashes and other system activity. \\
%PayPal &
  \textbf{[PayPal]} We collect the following types of user interaction data: app presentation,  {\color{red!60}binary, categorical, user input interactions} along with their  {\color{blue!60}frequency}. &
  \textbf{[PayPal]} When you visit our Sites, use our Services, or visit a third-party website for which we provide online Services, we and our business partners and vendors may use cookies and other tracking technologies to recognize you as a User and to customize your online experiences, the Services you use, and other online content and advertising; measure the effectiveness of promotions and perform analytics; and to mitigate risk, prevent potential fraud, and promote trust and safety across our Sites and Services. \\
%Duolingo &
  \textbf{[Duolingo]} We collect the following types of user interaction data: app presentation, binary, categorical, user input, gesture interactions, along with their {\color{blue!60}frequency} and duration. &
  \textbf{[Duolingo]} We do record the following data: Patterns, Clicks, Mouse movements, Scrolling, Typing, Pages visited, Referrers, URL parameters, Session duration. \\
%Amazon Prime Videos &
  \textbf{[Amazon Prime Videos]} We collect the following types of user interaction data: app presentation, binary, categorical, user input,  {\color{red!60}gesture interactions}, along with their  {\color{blue!60}frequency, duration and motion details}. &
  \textbf{[Amazon Prime Videos]} We automatically collect and store certain types of information about your use of Amazon Services including your interaction with content and services available through Amazon Services. List of examples: search for products or services in our stores and download, stream, view, or use content on a device, or through a service or application on a device. \\
%Yazio &
  \textbf{[Yazio]} We collect the following types of user interaction data: binary and  {\color{red!60}user input} interactions, along with their  {\color{blue!60}frequency}. &
  \textbf{[Yazio]} The Firebase Analytics service helps to determine the interactions of App users by recording, for instance, the first time the App is opened, deinstallations, updates, system crashes and how often the App is used. The service also records and analyses certain user interests. \\
%Fasion Famous &
  \textbf{[Fasion Famous]} We collect the following types of user interaction data:  {\color{red!60}app presentation, binary, user input, gesture and composite gesture interactions}, along with their  {\color{blue!60}frequency, duration and motion details}. &
  \textbf{[Fasion Famous]} Information that may be collected automatically: Data and analytics about your use of our Services. Data we collect with cookies and similar technologies: Data about your use of our Services, such as game interaction and usage metrics. \\
%Picsart &
  \textbf{[Picsart]} We collect the following types of user interaction data:  {\color{red!60}app presentation, binary, gesture and composite gesture interactions}, along with their  {\color{blue!60}frequency, duration and motion details}. &
  \textbf{[Picsart]} Our servers passively keep an electronic record of your interactions with our services, which we call ``log data''. We collect and combine data about the devices you use to access Picsart, and data about your device usage and activity. \\
%Dezor &
  \textbf{[Dezor]} We collect the following types of user interaction data: app presentation, binary,  {\color{red!60}categorical, user input interactions}, along with their  {\color{blue!60}frequency}. &
  \textbf{[Dezor]} The information collected by log files include internet protocol (IP) addresses, browser type, Internet Service Provider (ISP), date and time stamp, referring/exit pages, and possibly the number of clicks. \\\hline
\end{tabular}%
}
\end{table*}

%https://www.yazio.com/en/privacy
%https://www.booking.com/content/privacy.en-gb.html
%https://www.panteon.games/en/privacy-policy/
%https://picsart.com/privacy-policy
%https://eur.shein.com/Privacy-Security-Policy-a-282.html
%https://www.paypal.com/c2/legalhub/privacy-full
%https://www.duolingo.com/privacy
%https://www.amazon.co.uk/gp/help/customer/display.html?nodeId=201909010&pop-up=1
%https://www.tiktok.com/legal/page/eea/privacy-policy/en
%https://www.dezor.net/privacy

\subsection{Threats to Validity}
%While our study provides valuable insights into user interaction data collection practices in mobile apps, there are limitations to our methodology that should be taken into consideration when interpreting the results. 
Firstly, our analysis is based on the invocation of the top 20 analytics services, so if an app uses a service outside of this list or develops its own analytics service from scratch, we will not detect it.
%may not be able to identify such collections. 
Secondly, it is worth noting that our study is limited to analyzing data collection practices in Android apps. The information regarding claims on data types and means of collection were derived from the documentation of Android and related analytics libraries.
%We did not perform a massive user survey to capture additional possibilities, which could have provided further insights into user interaction data collection practices.

Another limitation is that the APP-350 corpus is three years old, and privacy policies may have since changed. 
A more recent policy corpus could have led to new insights about current practices in user interaction data collection. 
Additionally, we only manually fact-checked 10 apps due to the complexity of fitting different UI types and callbacks into the fixed six data types and three means of collection. 
%A more extensive machine learning model could be developed to automate this process and check a larger number of apps for consistency in privacy policies and actual data collection practices.

\section{Related Work}
The related work can be categorized into two themes: code analysis, privacy policy analysis, policy generation, and the privacy implications of analytics services.
Various studies have examined the extent to which information can be extracted from privacy policies~\cite{liu2018towards,ramanath2014unsupervised}, particularly related to opt-out mechanisms~\cite{sathyendra2017identifying}, purposes for app permission usages~\cite{baalous2018dangerous}, and sections relevant under the GDPR~\cite{tesfay2018read}. PrivacyFlash Pro~\cite{Zimmeck2021PrivacyFlashPA} was proposed as a privacy policy generator for iOS apps that leverages static analysis to identify privacy-related code features.

Other studies have focused on how analytics services can be used to capture user data. 
Alde~\cite{8660581} proposed a solution that uses both static and dynamic analysis to detect the key information analytics libraries collect, mostly on device-level information. 
On the other hand, PAMDroid~\cite{9284006} identifies personal data that flows into analytics services and defines it as a misconfiguration.

These studies have contributed to the understanding of privacy policies and data collection practices in mobile apps. 
However, there is a lack of research specifically on the practices of user interaction data collection and the transparency of related claims in privacy policies. That is the subject of this paper.

\section{Conclusion and Future Work}
This paper addresses the lack of transparency in user interaction data collection practices in mobile apps. 
We proposed a standardized claim template for user interaction data collection and utilized static analysis to extract evidence of these claims from Android apps. 
Our analysis of 100 popular apps revealed that 89\% of them collected user interaction data using various types and means. 
However, when we manually fact-checked the top 10 apps, we found incompleteness in all of their claims about such collection.
This highlights the importance of clear and comprehensive claims related to user interaction data collection and the potential of standardized claims to increase transparency. 
%Our contributions include the proposal of a standardized claim for user interaction data collection, which has the potential to increase transparency and accountability for mobile apps. 
%Additionally, we provided valuable insights into the types and means of user interaction data collection in popular apps, as well as the completeness and consistency of claims made in privacy policies.

%Despite the valuable insights provided by our study, 
There are limitations to our approach that point toward future work: 
Our analysis only included the top 20 analytics services, and our study was limited to Android apps. 
Moreover, our manual fact-checking of the 10 top apps relied on our interpretation of their policies. 
To address these limitations, future research could employ machine learning models to further automate the fact-checking process and get more comprehensive results.

In conclusion, our study underscores the significance of complete claims related to user interaction data collection in mobile apps. 
The findings of our study provide a foundation for future research and policy efforts aimed at increasing transparency in mobile app data collection practices.

\section*{Acknowledgements}
This work is part of the Privacy Matters (PriMa) project. 
The PriMa project has received funding from European Union’s Horizon 2020 research and innovation program under the Marie Skłodowska-Curie grant agreement No. 860315.

 \bibliographystyle{splncs04}
 \bibliography{mybibliography}

\begin{thebibliography}{10}
\providecommand{\url}[1]{\texttt{#1}}
\providecommand{\urlprefix}{URL }
\providecommand{\doi}[1]{https://doi.org/#1}

\bibitem{Androida43:online}
AppTornado: {AppBrain}: Android analytics libraries.
  \url{https://www.appbrain.com/stats/libraries/tag/analytics/android-analytics-libraries},
  (Accessed on 03/04/2023)

\bibitem{arzt2014flowdroid}
Arzt, S., Rasthofer, S., Fritz, C., Bodden, E., Bartel, A., Klein, J.,
  Le~Traon, Y., Octeau, D., McDaniel, P.: Flowdroid: Precise context, flow,
  field, object-sensitive and lifecycle-aware taint analysis for {A}ndroid
  apps. ACM SIGPLAN Notices  \textbf{49}(6),  259--269 (2014)

\bibitem{baalous2018dangerous}
Baalous, R., Poet, R.: How dangerous permissions are described in android apps'
  privacy policies? In: Proceedings of the 11th International Conference on
  Security of Information and Networks. pp.~1--2 (2018)

\bibitem{bird2009natural}
Bird, S., Klein, E., Loper, E.: Natural language processing with Python:
  analyzing text with the natural language toolkit. O'Reilly Media, Inc. (2009)

\bibitem{7736910}
Buriro, A., Akhtar, Z., Crispo, B., Del~Frari, F.: Age, gender and
  operating-hand estimation on smart mobile devices. In: 2016 International
  Conference of the Biometrics Special Interest Group (BIOSIG). pp.~1--5 (2016)

\bibitem{Cysneiros2009AnIA}
Cysneiros, L.M., Werneck, V.: An initial analysis on how software transparency
  and trust influence each other. In: Workshop em Engenharia de Requisitos
  (2009)

\bibitem{dumais2014understanding}
Dumais, S., Jeffries, R., Russell, D.M., Tang, D., Teevan, J.: Understanding
  user behavior through log data and analysis. Ways of Knowing in HCI pp.
  349--372 (2014)

\bibitem{fischer2016transparency}
Fischer-H{\"u}bner, S., Angulo, J., Karegar, F., Pulls, T.: Transparency,
  privacy and trust--technology for tracking and controlling my data
  disclosures: Does this work? In: Trust Management X: 10th IFIP WG 11.11
  Conference, IFIPTM 2016, Darmstadt, Germany, July 18-22, 2016, Proceedings
  10. pp. 3--14. Springer (2016)

\bibitem{gadotti2022pool}
Gadotti, A., Houssiau, F., Annamalai, M.S.M.S., de~Montjoye, Y.A.: Pool
  inference attacks on local differential privacy: Quantifying the privacy
  guarantees of apple's count mean sketch in practice. In: USENIX Security 22.
  pp. 501--518 (2022)

\bibitem{spacy2}
Honnibal, M., Montani, I.: {spaCy 2}: Natural language understanding with
  {B}loom embeddings, convolutional neural networks and incremental parsing
  (2017)

\bibitem{jain2019gender}
Jain, A., Kanhangad, V.: Gender recognition in smartphones using touchscreen
  gestures. Pattern Recognition Letters  \textbf{125},  604--611 (2019)

\bibitem{liu2018towards}
Liu, F., Wilson, S., Story, P., Zimmeck, S., Sadeh, N.: Towards automatic
  classification of privacy policy text. Carnegie Mellon University  (2018)

\bibitem{8660581}
Liu, X., Liu, J., Zhu, S., Wang, W., Zhang, X.: Privacy risk analysis and
  mitigation of analytics libraries in the android ecosystem. IEEE Transactions
  on Mobile Computing  \textbf{19}(5),  1184--1199 (2020).
  \doi{10.1109/TMC.2019.2903186}

\bibitem{morey2015customer}
Morey, T., Forbath, T., Schoop, A.: Customer data: Designing for transparency
  and trust. Harvard Business Review  \textbf{93}(5),  96--105 (2015)

\bibitem{ramanath2014unsupervised}
Ramanath, R., Liu, F., Sadeh, N., Smith, N.A.: Unsupervised alignment of
  privacy policies using hidden markov models. In: Proceedings of the 52nd
  Annual Meeting of the Association for Computational Linguistics. pp. 605--610
  (2014)

\bibitem{SeveralE79:online}
Rzhevkina, A.: {Several EU countries banned Google Analytics - here are some
  alternatives}.
  \url{https://www.contentgrip.com/eu-countries-ban-google-analytics/}
  (September 2022), (Accessed on 03/12/2023)

\bibitem{sathyendra2017identifying}
Sathyendra, K.M., Wilson, S., Schaub, F., Zimmeck, S., Sadeh, N.: Identifying
  the provision of choices in privacy policy text. In: Proceedings of the 2017
  Conference on Empirical Methods in Natural Language Processing. pp.
  2774--2779 (2017)

\bibitem{story2019natural}
Story, P., Zimmeck, S., Ravichander, A., Smullen, D., Wang, Z., Reidenberg, J.,
  Russell, N.C., Sadeh, N.: Natural language processing for mobile app privacy
  compliance. In: AAAI Spring Symposium on Privacy-Enhancing Artificial
  Intelligence and Language Technologies (2019)

\bibitem{tesfay2018read}
Tesfay, W.B., Hofmann, P., Nakamura, T., Kiyomoto, S., Serna, J.: {I read but
  don't agree: Privacy policy benchmarking using machine learning and the EU
  GDPR}. In: Companion Proceedings of the The Web Conference 2018. pp. 163--166
  (2018)

\bibitem{verkasalo2010analysis}
Verkasalo, H.: Analysis of smartphone user behavior. In: 2010 Ninth
  International Conference on Mobile Business and 2010 Ninth Global Mobility
  Roundtable (ICMB-GMR). pp. 258--263. IEEE (2010)

\bibitem{vorm2022integrating}
Vorm, E., Combs, D.J.: {Integrating Transparency, Trust, and Acceptance: The
  Intelligent Systems Technology Acceptance Model (ISTAM)}. International
  Journal of Human--Computer Interaction  \textbf{38}(18-20),  1828--1845
  (2022)

\bibitem{9284006}
Zhang, X., Wang, X., Slavin, R., Breaux, T., Niu, J.: How does misconfiguration
  of analytic services compromise mobile privacy? In: 2020 IEEE/ACM 42nd
  International Conference on Software Engineering (ICSE). pp. 1572--1583
  (2020)

\bibitem{Zimmeck2021PrivacyFlashPA}
Zimmeck, S., Goldstein, R., Baraka, D.: {PrivacyFlash Pro}: Automating privacy
  policy generation for mobile apps. Proceedings 2021 NDSS Symposium  (2021)

\end{thebibliography}
\end{document}